\documentclass[twocolumn]{revtex4}
\usepackage{amsmath,amssymb,epsfig,bm}
\def\be{\begin{eqnarray}}
\def\ee{\end{eqnarray}}

\newcommand{\vecw}{\mathbf w}
\newcommand{\vecj}{\mathbf j}
\newcommand{\vecv}{\mathbf v}

\newcommand{\vecf}{\mathbf f}

\def\beps{{\mbox{\boldmath $\epsilon$}}}
\def\bnabla{{\mbox{\boldmath $\nabla$}}}
\def\bOmega{{\mbox{\boldmath $\Omega$}}}
\def\bomega{{\mbox{\boldmath $\omega$}}}
\def\bsigma{{\mbox{\boldmath $\sigma$}}}

\begin{document}
\title{Tkachenko modes as sources of quasiperiodic pulsar spin variations}
\author{Jorge Noronha}
\email{noronha@fias.uni-frankfurt.de}
\affiliation{Frankfurt Institute for Advanced Studies,
J. W. Goethe-Universit\"at,
D-60054 Frankfurt am Main, Germany
}

\author{Armen Sedrakian}
\email{sedrakian@th.physik.uni-frankfurt.de}
\affiliation{Institute for Theoretical Physics, J. W. Goethe-Universit\"at,
D-60054 Frankfurt am Main, Germany
}
\begin{abstract}
We study the long wavelength shear modes (Tkachenko waves)
of triangular lattices of singly quantized vortices in neutron
star interiors taking into account the mutual friction between
the superfluid and the normal fluid as well as the shear viscosity of the
normal fluid. The set of Tkachenko modes that propagate in
the plane orthogonal to the spin vector are weakly damped if the
coupling between the superfluid and the normal fluid is small. In strong coupling, their oscillation frequencies are lower
and are undamped for small and moderate shear viscosities.
The  periods of these modes are consistent with the
observed $\sim 100-1000$ day variations of spin for PSR 1828-11.
\end{abstract}

\maketitle

\section{Introduction}
\label{sec:intro}

During the last decade mounting evidence has emerged for the
existence of long-period oscillations in a handful of
pulsars~\cite{cordes,dalessandro,stairs00,hobbsthesis,hobbs}. An outstanding example
of this kind of phenomenon is observed in PSR 1828-11~\cite{stairs00}.
Its timing residuals are modulated with periods of 256 and 511 days, while a 1009-day periodicity is inferred with lower confidence.
These timing residuals coincide with periodic modulations
of the pulse shape, which is a strong indication for the
precessional motion(s) of the pulsar with the periods quoted
above~\cite{stairs00}. In fact, various models of precessing neutron
stars~\cite{Akgun:2005nd,Wasserman:2002ec,Cutler:2002np,Link:2001zr,Jones:2000ud}
fit the timing data fairly well. However, it is known that free precession
is incompatible with the existence of a superfluid in the pulsar's interior if the superfluid is strongly coupled to the normal fluid~\cite{Shaham77,Sedrakian:1998vi,Link:2003hq,Link:2006nc}. The mutual friction for the superfluids present in the crust and core of neutron stars, which are derived
using microscopic calculations, cover a broad range of values.
A reliable calculation necessarily involves the
superconducting and superfluid properties of the fluid(s) and
the non-superconducting material, which resists to vortex
motion. While precession remains a viable model for the quasiperiodic
oscillations observed in pulsar timing data, here we follow a different
route~\cite{Sedrakian:2004yq} by exploring the propagation of Tkachenko modes
in pulsars as the source of long term variations.

Charge neutral superfluids in neutron star interiors rotate
by forming an array of singly quantized vortices.
In their lowest energy state the vortices form a two-dimensional
triangular lattice. The lattice supports collective elastic modes,
Tkachenko waves, in which the vortices are displaced parallel to
each other~\cite{TKACHENKO1,TKACHENKO2,TKACHENKO3,TKACHENKO4}.
Their undamped propagation would lead to variations of the angular momentum
of the superfluid due to the local
variations of the density of the vortex lines as well as periodic variations
in the rotation and spin-down speeds of the star.

Ruderman \cite{RUDERMAN} pointed out that the frequency of the
Tkachenko modes is of the order of several hundred days and
these modes could be responsible for the quasiperiodic timing
residuals observed in the Crab pulsar.
Little attention has been paid to the role played by the Tkachenko
modes in neutron stars since Ruderman's 1970 paper.
Here we reinvestigate the propagation of
Tkachenko modes in neutron star superfluids
and calculate how these modes are damped by mutual friction and
the shear viscosity of normal matter. These factors are clearly
important for the continuous propagation of these modes, which would certainly lead to observable effects.
Provided that the observed variations are
indeed caused by the shear modes of the lattice we can ask
the question: what do the quasi-sinusoidal variations tell us
about the microscopic physics governing superfluids in neutron stars?

This question can be answered by describing the physics
present at micro- and mesoscopic scales (the upper limit on
these scales is set by the size of the neutron vortex) in terms of
the few parameters that enter the equations of superfluid
hydrodynamics. These parameters include the kinetic coefficients
and other local characteristics (e.~g., density, baryon and lepton
fractions, etc). The predictions obtained within this hydrodynamic
description can be presented in a form that is independent of the details
of physics at the micro- and mesoscopic scales.

In this paper we pursue this ``model independent'' approach, although some comments about the physics present at intermediate scales are in order. A key problem concerning the crust of neutron stars
is the question of whether neutron vortices are pinned to the
crustal nuclei both under static and dynamical conditions.
The static problem always has pinning solutions for infinitesimally
attractive interactions between a neutron vortex and a nucleus
(for repulsive interactions the vortex is localized in between
the lattice sites)~\cite{Donati,Avogadro}. The dynamical
problem of repinning under the action of external
superflow~\cite{Sedrakian_repinning} allows for repinning solutions
only for strong pinning potentials and current estimates
for the pinning energy do not favor repinning solutions. However, if the
axisymmetry of the problem is lost, as in the case of precessing
neutron stars, the vortices may not pin at all~\cite{Link:2001dg}.

Another issue concerning neutron star crusts
is that they may often undergo starquakes~\cite{Franco:1999ap,Middleditch:2006}, which drive the vortex lattice out of its
equilibrium state. In fact, the form of the lattice may change from a
triangular shape into some other structure reflecting the symmetries
of the underlying nuclear lattice. Moreover, other possible scenarios include cases where the vortices may acquire multiple quanta of circulation or the lattice may contain impurities
and vacancies. Thus, our working assumption that the ground state of the lattice consists of a simple triangular array of vortices may not be applicable to the crusts of frequently quaking neutron stars.

Regarding the cores of neutron stars, a major question is whether the proton fluid
is a type-I~\cite{Charbonneau:2007db,Alford:2005ku,Sedrakian:1997gg} or type-II
superconductor~\cite{Baym69,Baym:1975mf,muzikar:1981,MFLUIDS1,
MFLUIDS2,MFLUIDS3,Mendell:1997fu,MFLUIDS4,Sedrakian:1998ki,MFLUIDS5,Akgun:2007ph}
(it may also become unpaired
at high densities). The friction in the case of type-II superconductors
is large~\cite{Sedrakian:1998ki}, which implies that
precession is impossible. If, however, the proton fluid is a
type-I superconductor the friction is found to be small and
compatible with precession~\cite{Sedrakian:2004yq}.

Finally, we would like to point our that the
superfluids may develop quantum turbulence when
the velocity difference between the superfluid and the normal  fluid
exceeds a certain critical value. Turbulent states in superfluid He
may be generated in many ways, e.~g., by driving the normal fluid
along the vortex lines through a temperature gradient.
Neutron stars superfluids may also develop turbulent
states~\cite{Peralta:2005xw,Peralta:2006um,Andersson:2007uv},
which could be caused either by unstable
precession~\cite{Glampedakis:2007hx} or tectonic activity in
the crusts~\cite{Melatos:2007px}.

This paper is organized as follows. In Sec.~\ref{sec:hydro} we
recapitulate the equations of superfluid hydrodynamics that include
the combined effects of vortex tension, mutual friction, and
shear viscosity. Sec.~\ref{sec:modes} is devoted to the
derivation of the characteristic equation for the Tkachenko
and inertial modes and their numerical study. Our conclusions are
summarized in Sec.~\ref{sec:conclusions}.

\section{Superfluid hydrodynamics with vortex tension}
\label{sec:hydro}

It is convenient at this point to discuss the three main length scales that appear in our study in more detail. The relevant length scales are the vortex core radius
$\sim 10^{-12}$~cm, the intervortex spacing $\sim 10^{-3}$~cm, and
the size of the superfluid phase $\sim 10^5$~cm (we refer to these three different scales as micro, meso, and macroscopic scales). A hydrodynamic description requires averaging over
the mesoscopic scales. In order to describe the deformations of the vortex lattice,
one needs an additional dynamical variable, the local
deformation of the vortex lattice $\beps ({\bf r})$, which can be
incorporated into the Bekarevich-Khalatnikov superfluid
hydrodynamics~\cite{BK_HYDRO}. A hydrodynamic description
of superfluids that includes lattice deformations has
been studied by a number of authors~\cite{VOLOVIK_DOTSENKO,BAYM_CHANDLER,CHANDLER_BAYM,Reisenegger} and
in this work we use the Baym-Chandler version of superfluid
hydrodynamics~\cite{CHANDLER_BAYM} to study the Tkachenko
modes and their damping due to mutual friction and shear
viscosity.

The fluid motions are naturally separated into the center of mass
motion (known as first sound) and second sound, which corresponds to temperature dependent relative oscillations between the superfluid and the normal component. These motions are
conveniently described in terms of the total mass current
${\bf j} = \rho_N {\bf v}_N +\rho_S {\bf v}_S$ and the relative
velocity ${\bf w} = {\bf v}_N- {\bf v}_S$, where ${\rho}_S$ (${\rho}_N$) and
${\bf v}_S$ (${\bf v}_N$) are the superfluid (normal fluid) density and velocity, respectively. In addition, one needs an equation for the time variations of the
lattice deformation $\beps ({\bf r})$. We note in passing that $\rho=\rho_N+\rho_S$ is the total mass density and the tiny effects arising due to the inertia of the vortex lines are neglected.

The linearized version of the fundamental superfluid hydrodynamic
equations written for the net mass current, the relative velocity,
and the superfluid velocity are
\be
\label{eq:1}
\frac{\partial {\vecj}}{\partial t}
+\left(2{\bOmega}\times {\vecj}\right) + {\bf C}
+\bsigma+\bnabla P+\rho \bnabla \phi &=&0,\\
\label{eq:2}
\frac{\partial {\vecw}}{\partial t}
+\left(2{\bOmega}\times {\vecw}\right)
-\frac{\bsigma}{\rho_S}\ -{\vecf} &=& 0,\\
\label{eq:3}
\frac{\partial {\vecv}_S}{\partial t}+
\left(2 {\bOmega}\times \frac{\partial {\beps} }{\partial t} \right)
+ \frac{\bnabla P}{\rho} +\bnabla\phi &=& 0,
\ee
where ${\bOmega}=(0,0,\Omega)$ is the spin vector ($\Omega$ is the pulsar rotation frequency),
$P = P_0-\rho ({\bOmega}\times {\bf r})^2/2$,  $P_0$ is the
pressure in the fluid at rest, and $\bsigma$ is the vortex elastic
force density defined as
\be
\bsigma = \mu_S \left[
2 \bnabla_{\perp} \cdot (\bnabla_{\perp}\cdot \beps)-\bnabla_{\perp}^{2}\beps\right]
- 2 \Omega \lambda \frac{\partial^2\beps}{\partial z^2},
\ee
where $\bnabla_{\perp}$ is the gradient in the $x-y$ plane and $\mu_S = \rho_S \hbar\Omega/8m_N$ is the shear modulus of the triangular vortex lattice calculated by Tkachenko~\cite{TKACHENKO1,TKACHENKO2,TKACHENKO3,TKACHENKO4}. Moreover, $m_N$ is the bare neutron mass and the vortex tension is given by
\be
\lambda = \frac{\hbar\rho_S}{8m_N}{\rm ln}\left(\frac{b}{a}\right),
\ee
where $a$ is the coherence length and $b = (\pi\hbar/\sqrt{3}m_N\Omega)$ is
the vortex radius of the triangular lattice. The Newtonian
gravitational potential $\phi$ satisfies the equation
\be
\nabla^2\phi = \nabla^2
(\phi_S+\phi_N) = 4\pi G (\rho_S+\rho_N),
\ee
where $G$ is the Newton's constant and $\phi_{S}$ and $\phi_{N}$
are the gravitational potentials of the superfluid and the normal
fluid, respectively. The force density ${\bf C}$ is defined as $C_{i}=\nabla_{k}\tau_{ik}$, where $\tau_{ik}$ is the viscous stress tensor
\be
\tau_{ik} = -\eta \left(\nabla_i v_{Nk} +\nabla_k v_{Ni} -\frac{2}{3}\delta_{ik}
\bnabla\cdot{\bf v}_N\right),
\ee
whereas $\eta$ is the shear viscosity. Note that we do not take into account the effects from bulk viscosity and thermal conductivity. Finally, the mutual friction force is
\be\label{mf}
{\bf f} &=& \beta\rho_S \left[
{\bf n}\times \left[\bomega\times\left(\frac{\partial {\beps} }{\partial
t}-{\bf v}_N\right)\right]\right]\nonumber\\
&&\hspace{2cm}+\beta'\rho_S\left[\bomega\times\left(\frac{\partial {\beps}}{\partial t}
-{\bf v}_N\right)\right],
\ee
where $\bomega=\bnabla\times {\bf v}_S $ is the quantum circulation vector, ${\bf n}\equiv \bomega/\omega$, and $\beta$ and $\beta'$ are the phenomenological mutual friction coefficients.

The dissipative terms in the hydrodynamic equations defined above (such as the
mutual friction forces) are the most general expressions that can be used in the description of a rectilinear vortex lattice in equilibrium that are still compatible with the conservation
laws and the assumption that the dissipative function is a positive
quadratic form of the perturbations (higher order terms are neglected).
Furthermore, we would like to remark that Baym-Chandler hydrodynamics only describes the linear order corrections in the lattice displacements, which are considered to be small. The neutron vortex lattice in neutron stars may, however, become unstable towards forming a tangle with turbulent superfluid flow, which would require a thorough revision of the hydrodynamic equations.
In particular, the form of the mutual friction force (\ref{mf}) should
be changed to the one suggested by Groter and Mellink~\cite{Groter:1949} where ${\bf f}\propto {\bf w}$. These aspects of superfluid
dynamics are beyond the scope of this work (see Refs.~\cite{Peralta:2005xw,Peralta:2006um,Andersson:2007uv}).

\section{Oscillation modes}
\label{sec:modes}

We consider plane wave perturbations with respect to the equilibrium,
which corresponds to uniform rotation. We use a Cartesian system
of coordinates where the $z$-axis is directed along the spin vector $\bomega$.
The vectors ${\bf j}$ and ${\bf w}$ can be decomposed
into transverse and longitudinal parts, i.e., ${\bf j}={\bf j}_{t}+{\bf j}_{l}$ and ${\bf w}={\bf w}_{t}+{\bf w}_{l}$. The transverse parts we are interested in satisfy the condition
\be
\bnabla \cdot {\vecj}_t =\bnabla \cdot {\bf w}_t= 0.
\ee
The perturbation equation for the transverse components of the
vectors ${\bf j}$ and ${\bf w}$ derived from Eqs.
(\ref{eq:1})-(\ref{eq:3}) are (hereafter the subscript $t$ is suppressed)
\be \label{eq:10}
\frac{\partial j_i}{\partial t} +
(2\epsilon_{lmn}\Omega_m j_n +\sigma_l+k_m\tau_{lm}) P_{il} &=& 0,\\
\label{eq:11}
\frac{\partial w_i}{\partial t} +
(2\epsilon_{lmn}\Omega_m w_n -\frac{\sigma_l}{\rho_S} - f_l) P_{il}
&=& 0,\\
\label{eq:12}
\frac{\partial v_i}{\partial t} +
2\epsilon_{lmn}\Omega_m \frac{\partial\epsilon_n}{\partial t} P_{il} &=& 0.
\ee
In the equations above we used the projector $P_{il} = \delta_{il}-k_ik_l/k^2$
where ${\bf k}$ is the wave vector. Our coordinate system is such that the wave vector lies on
the $z-x$ plane, i.e., ${\bf k } = (k\sin~\theta,  ~0, \,
k\cos~\theta)$ where $\theta$ is the angle formed by the vectors
$\bOmega$ and ${\bf k}$.

Writing the time perturbations as $j_i(t) \sim j_i~e^{2\Omega pt}$ (a similar definition is used for the other vectors) we obtain, after some algebra,
the characteristic equation ${\rm det}~
\vert\vert K_{ij}\vert\vert = 0$ where
\begin{widetext}
\be\label{MATRIX}
K_{ij}=\left(
\begin{array}{cccc}
p-\tilde\eta  \alpha d& (\gamma_{S}h-1) &
-\tilde\eta  \gamma_S \alpha d& -\gamma_S\gamma_N h\\
d+\gamma_{S}g & p-\tilde\eta & \gamma_S\gamma_N g& -\tilde\eta \gamma_S \\
-\hat\beta g & -\hat\beta^* h &
p+\hat\beta (d+\gamma_{N}g)& -\hat\beta^* (1-\gamma_{N}h)\\
-\hat\beta^* g & \hat\beta h &
\hat\beta^*(d+\gamma_{N}g) &p+\hat\beta (1-\gamma_{N}h)
\end{array}
\right).\nonumber\\
\ee
\end{widetext}
We used the following shorthand notations in the definition of the matrix $K_{ij}$: $\gamma_{N/S} = \rho_{N/S}/\rho$,
$d^{1/2} = \cos~\theta$, $\hat\beta^* = 1-\hat\beta'$, $\tilde\eta =
\eta k^2/(2\Omega\rho)$, $\alpha = (4-d)/3$, $\hat\beta = \gamma_N^{-1}
\beta$, and $\hat\beta' = \gamma_N^{-1}\beta'$. Moreover, we defined
\be
g &=& \frac{k^2}{4\Omega^2\rho_S}\left[\mu_S
-d (\mu_S-2\Omega\lambda) \right],\\
h &=& \frac{k^2}{4\Omega^2\rho_S}\left[\mu_S -
d(\mu_S+2\Omega\lambda) \right].
\ee
Note that the coefficients $g$ and $h$ are independent of the density
because $\mu_S\sim\rho_S$ and $\lambda\sim\rho_S$. The density appears only
through the normalization of the shear viscosity $\tilde \eta\sim \rho^{-1}$. The eigenmodes of the matrix in Eq.~(\ref{MATRIX}) provide the oscillatory modes in the general
case where the shear viscosity of normal matter and the mutual friction
are included. In the non-dissipative limit ($\beta= \beta'=\eta=0$)
the modes separate into two independent sets that describe the inertial and Tkachenko modes, respectively. The (real) eigenfrequencies of these modes in units of $2\Omega$ are
\be
p_I &=& \pm i\, d^{1/2},\quad p_T =\pm i\left[(d+g)\, (1-h)\right]^{1/2},
\ee
where the indices $I$ and $T$ refer to inertial and Tkachenko modes, respectively.

If the Tkachenko modes are generated within superfluid shells with the width of $R_c \sim 10$ km their corresponding wave vectors are of the order of
$k_{\rm min} = 2\pi/R_c \sim 6.28\times 10^{-6}$ cm, which sets the lower
limit on the wave vector. Since the hydrodynamic description breaks down at
length scales $\sim 10 b$ ($b\sim 10^{-3}$ cm is the intervortex
distance) the wave vector is bounded from above by the value
$k_{\rm max} = 73.3$ cm$^{-1}$. We are interested in the
small wave vector limit $k\sim k_{\rm min}$ that describes
vortex density waves across the entire superfluid shell.

The parameters $g$ and $h$ are of the order of $s = (\hbar^2k_{\rm min}^2/2m_n)
(8\hbar\Omega)^{-1}$. For instance, when $k = k_{\rm min}$ and $\Omega_{\star}
= 15.51$ Hz (the rotation frequency of  PSR 1828-11) we obtain that $s=10^{-14}$.
Therefore, for $s\ll 1$ the eigenmodes corresponding to Tkachenko
waves in the dissipationless limit are given by
$p_T =\pm i\left[(d+g)\right]^{1/2}$. In the limit $d\ll g$ where
the wave vectors are highly collinear to the spin vector we obtain that
$p_T = \pm i\sqrt{g}$ and, in the opposite limit $d\gg g$, the Tkachenko modes
become identical to the inertial modes, i.e., $p_T = p_I$.
\begin{figure}[t] 
\begin{center}
\includegraphics[height=7cm,width=8cm]{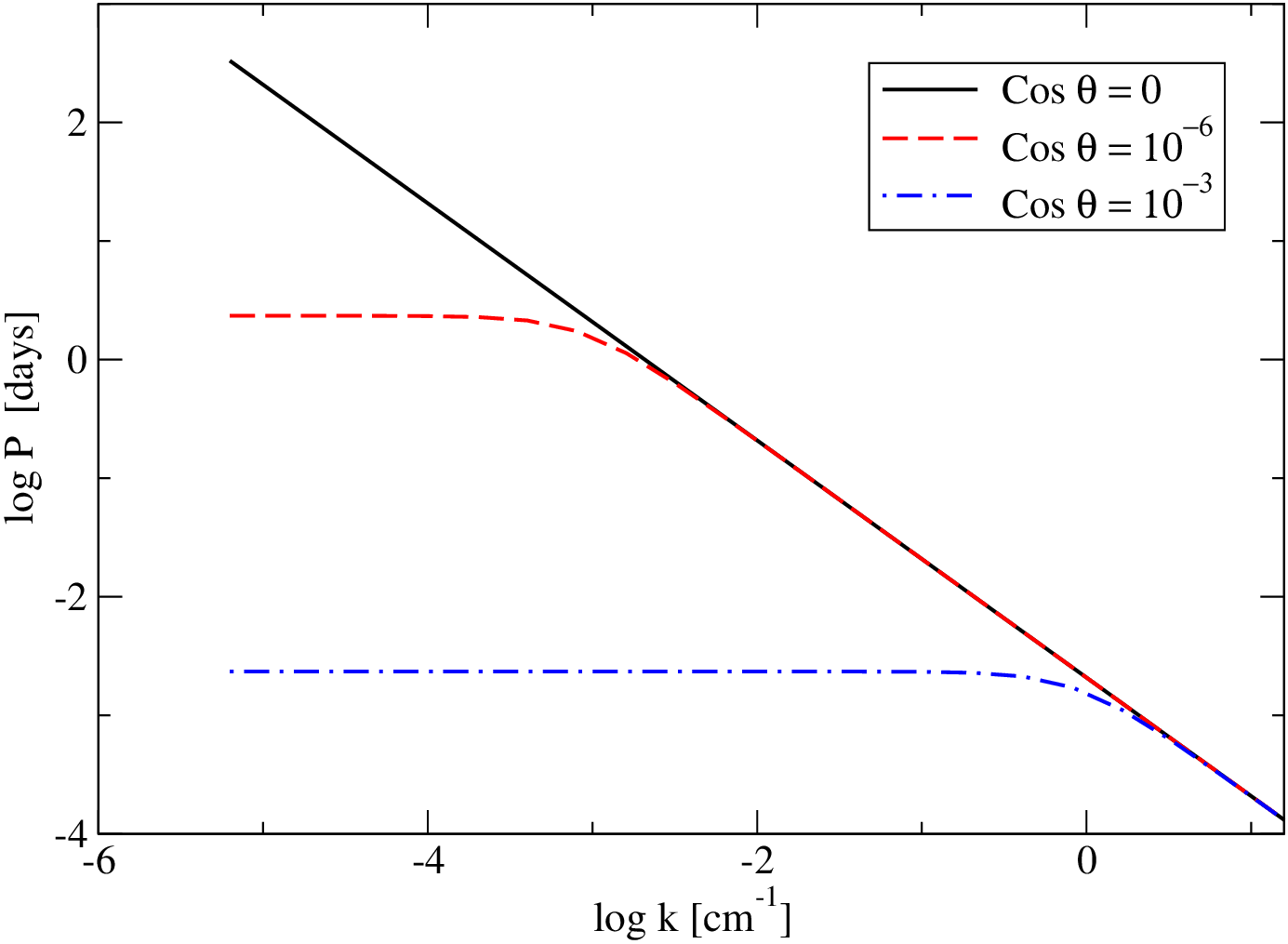}
\end{center}
\caption{(Color online) Dependence of the period
$P = 2\pi/\vert p_T\vert$ of the Tkachenko modes
on the wave vector for $d=0$ (solid black line),  $d=10^{-12}$
(dashed red line), and $d=10^{-6}$ (dashed-dotted blue line). For large wavelengths the periods are on the order of 100 days. In particular, for $\Omega_{\star} = 15.51$ Hz we obtain that
$P(k_{\rm min},d=0)  = 331$ days if the core size is
$R_c = 10$ km and 256 days when $R_c=7.7$ km.}
\label{MSfig:fig1}
\end{figure}

Fig.~\ref{MSfig:fig1} displays the period of the Tkachenko
modes without dissipation as a function of their wave vector. Only the long wavelength perturbations have periods on the order of 100 days, which are then relevant for observations. In this limit the periods rapidly decrease for perturbations with finite $d$.
The period $P(k_{\rm min},d=0)  = 331$ days for $\Omega_{\star} = 15.51$ Hz
suggests that the shortest of the periods observed in PSR 1828-11, which corresponds to 256 days,
should be identified with the fundamental oscillation mode. Oscillations with larger periods should then be identified with the higher-order harmonics of this mode. A period of 256 days can be obtained by adopting $R_c = 7.7$ km, which translates into $k = 8.16\times 10^{-6}$ cm$^{-1}$. This value is close to the upper limit for the size of any superfluid region inside a medium-heavy neutron star. It is important to remark that if the width of the superfluid region where Tkachenko waves can be found is at least one order of magnitude (or more) smaller than the values of $R_c$ quoted
above the modes will be too fast to account for the long-period oscillations observed in PSR 1828-11 (as long as the renormalization effects due to mutual friction discussed below are neglected).

We now consider the effects of the shear viscosity and the mutual friction on the propagation of Tkachenko modes. It is convenient to use the drag-to-lift ratios $\zeta$ and $\zeta'$ instead
of $\beta$ and $\beta'$ to describe mutual friction. These ratios are related by the following equations
\be
\beta = \frac{\zeta}{[(1-\zeta')^2+\zeta^2]},\quad
\beta' = 1- \frac{\beta(1-\zeta')}{\zeta}.
\ee
Microscopic calculations indicate that $\zeta'\simeq 0$.
The limit $\zeta\to 0$ corresponds to weak coupling
between the vortices and the normal fluid, while $\zeta\to \infty$ implies strong coupling.

\begin{figure}[tb] 
\begin{center}
\vskip 1cm
\includegraphics[height=8.5cm,width=7.5cm]{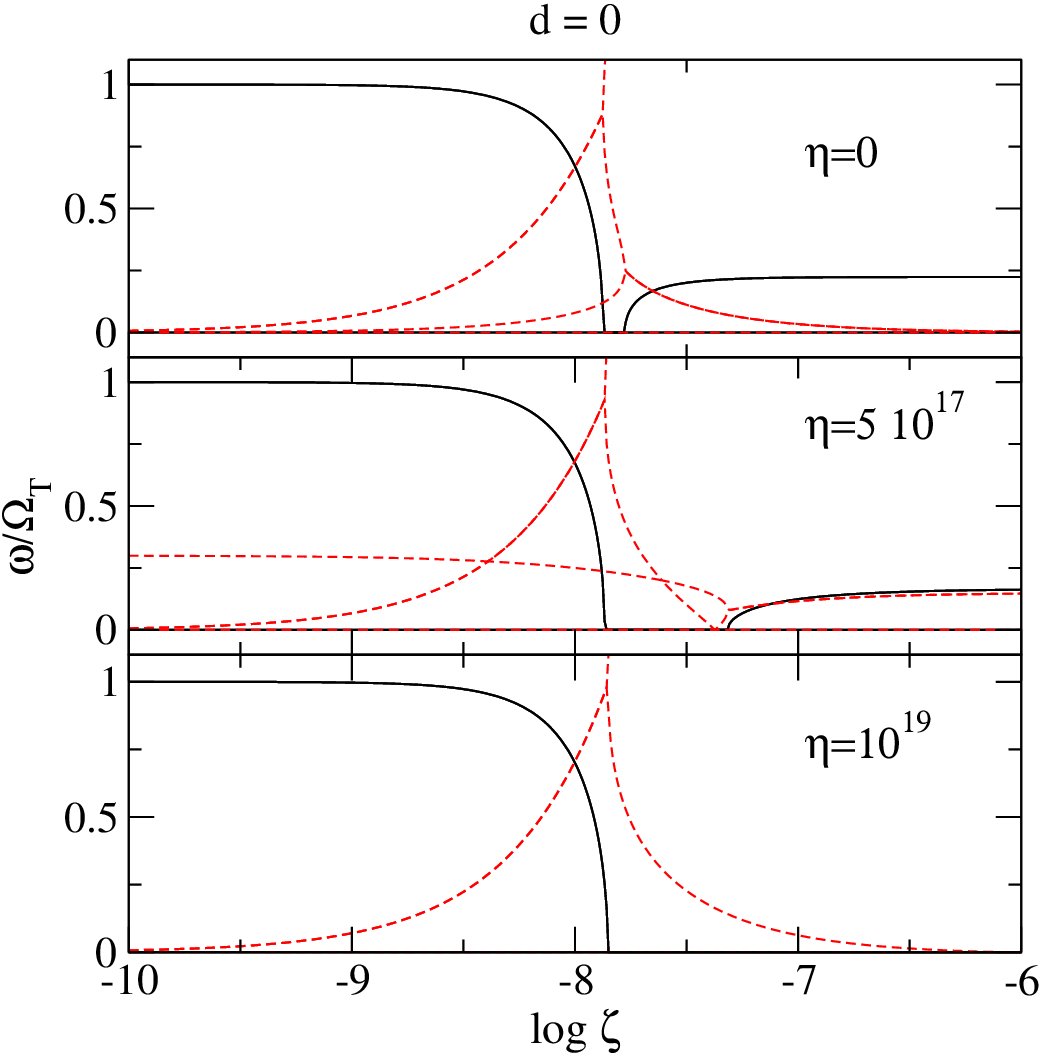}
\end{center}
\caption{(Color online)
Dependence of the real (solid black line) and imaginary
(dashed red line) parts of the Tkachenko modes ($\omega = ip$)
on the drag-to-lift ratio
$\zeta$ for $\eta = 0$ (upper panel), $\eta = 7.5\times 10^{17}$ (middle
panel), and $\eta = 1\times 10^{19}$ (lower panel) in
dyn s cm$^{-2}$ units. All modes are normalized
by the non-dissipative value of the Tkachenko mode $\Omega_T = 2\pi/P$.
The modes are computed taking $d=0$, which means that there are no inertial modes.
}
\label{MSfig:fig2}
\end{figure}

Figure \ref{MSfig:fig2} shows the dependence of the modes derived from
Eq.~(\ref{MATRIX}) on the drag-to-lift ratio $\zeta$ for several values of the shear viscosity and $d=0$. The value of $\tilde\eta$ is determined assuming a constant density of $3\times 10^{14}$ g cm$^{-3}$. In the limit where $\zeta$ and $\eta$ vanish we recover the results for the
non-dissipative case discussed above. For $\eta = 0$
the real part of the Tkachenko mode, which is doubly degenerate,
vanishes only in a narrow window of values of $\zeta$. For larger values of $\zeta$, which corresponds to the strongly coupled region, the real part reaches an asymptotic value that is about 25\% smaller than its value in the undamped limit. Note that in our plots only the regions where the modes change significantly are shown.

Assuming that the normal fluid is inviscid, the results in Fig.\ \ref{MSfig:fig2} imply that
there are oscillations with even longer periods in the strongly coupled limit. The Tkachenko modes are significantly damped by mutual friction in the region where the real part vanishes. There the number of imaginary roots of the characteristic equation increases by one. Moreover, one of the imaginary roots is given by ${\rm Im}~\omega  = i\beta\Omega$, which continues beyond the
figure's $y$ scale. This reflects the damping of the differential rotation between the superfluid and the normal fluid caused by mutual friction. This damping has no effect on the Tkachenko modes in strong coupling. For moderate values of viscosity ($\eta = 5 \times 10^{17}$
dyn s cm$^{-2}$) the real part of the Tkachenko mode is reduced in the strongly coupled region. In this case, its imaginary part is smaller than the real part and, therefore, the oscillations are weakly damped. For large values of the shear viscosity ($\eta\sim 10^{19}$ dyn s cm$^{-2}$)
the real part of the Tkachenko mode vanishes in the strongly coupled limit. Finally, note that there are no inertial modes when $d = 0$.

\begin{figure}[!] 
\begin{center}
\includegraphics[height=8.5cm,width=7.5cm]{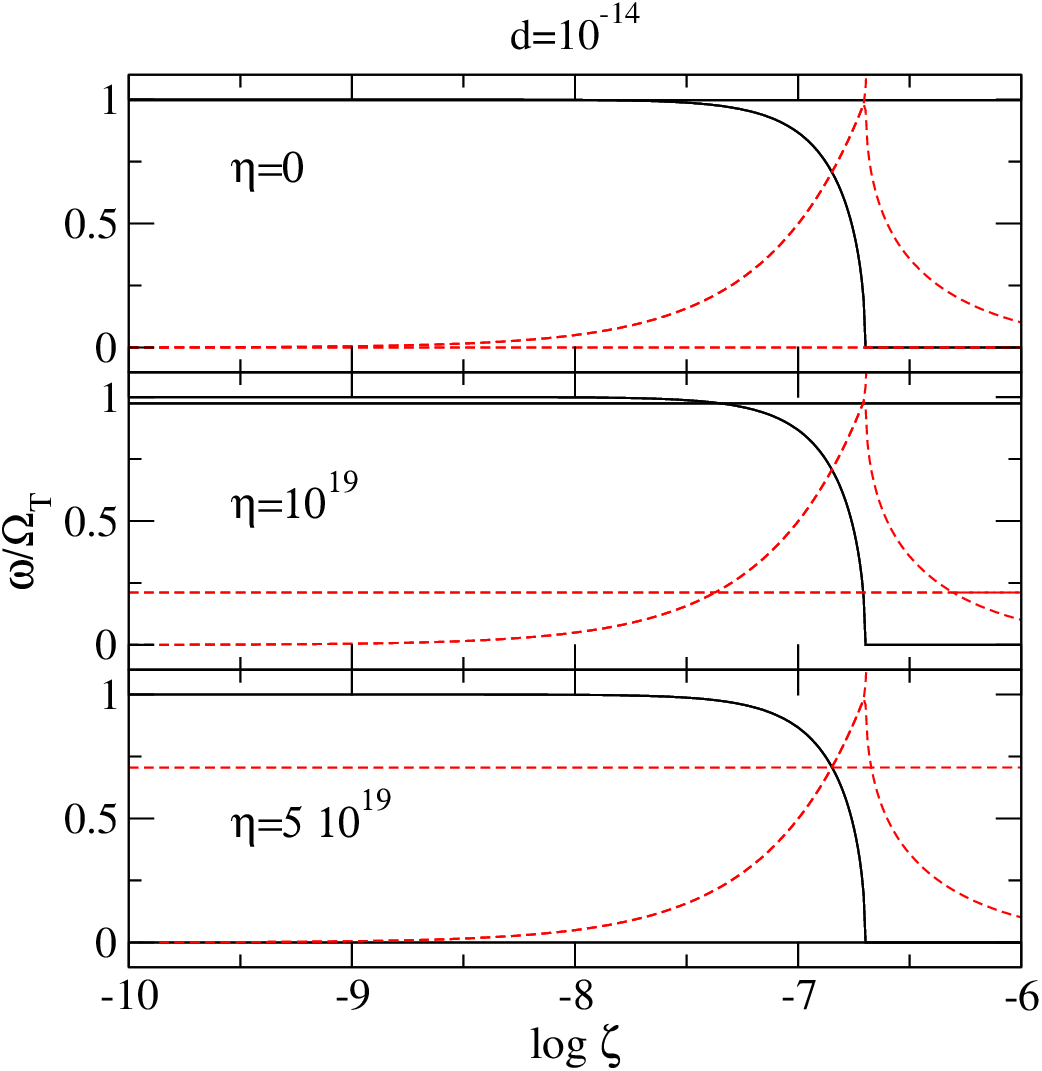}
\end{center}
\caption{(Color online) The same as in Fig.~\ref{MSfig:fig2} in case of $d=10^{-14}$ and $\eta =0$ (upper panel), $\eta = 10^{19}$
(middle panel), and $\eta = 5\times 10^{19}$ (lower panel) in
dyn s cm$^{-2}$ units. Note that both the Tkachenko and inertial modes are displayed here.
}
\label{MSfig:fig3}
\end{figure}

The modes when $d=10^{-14}$ are shown in
Fig.~\ref{MSfig:fig3}. As discussed above, in the non-dissipative limit
the Tkachenko and inertial modes coincide for sufficiently large $d$.
For $\eta = 0$ the modes can be distinguished in the strongly coupled limit because the Tkachenko mode vanishes for sufficiently large values of $\zeta$. When larger viscosities are considered ($\eta > 10^{19}$ dyn s cm$^{-2}$) the difference between the real parts of the
inertial and Tkachenko modes can be clearly resolved. If we increase $\eta$ even further we see that the real part of the inertial mode decreases and the imaginary part, which increases with $\eta$, becomes relevant. Finally, the real part of the inertial mode vanishes at $\eta \simeq 5 \times 10^{19}$ dyn s cm$^{-2}$.

The outer cores of neutron stars are mainly composed of light baryons,
which pair in the isospin triplet states, and leptons. For densities of $2-3\times 10^{14}$ g cm$^{-3}$ and temperatures of $T\sim 10^8$ K the shear viscosity of the electron fluid was determined to be in the interval between $8 - 40 \times 10^{17}$ dyn s cm$^{-2}$~\cite{FLOWERS_ITOH}. This value of the temperature is a realistic upper bound on the temperature in the core of neutron stars except for very young objects such as the Vela and Crab pulsars. For colder stars the viscosity could
be a few orders of magnitude larger because $\eta \sim T^{-2}$.

\section{Conclusions}
\label{sec:conclusions}

Our results indicate that Tkachenko modes are broadly consistent
with the weakly coupled theories between the superfluid and the normal
fluid, independent of the value the shear viscosity. The subclass
with $d=\cos^2\theta=0$ has periods that are consistent with
the lowest observed periodicity in PSR 1828-11 of 256 days.

The existence of Tkachenko modes in the strongly coupled region depends on the shear
viscosity of normal matter. For low viscosities the Tkachenko
modes are (in strong coupling) renormalized to values that are
a few times smaller than their non-dissipative limits. This implies
that in strong coupling the Tkachenko oscillations have periods that are larger
than their non-dissipative counterparts. In fact, the damping caused
by mutual friction is not always strong enough to preclude an
oscillatory behavior. Therefore, we conclude that the long term variation
in the spin of PSR 1828-11 can in principle be explained in terms
of Tkachenko oscillations within superfluid shells for a broad range
of values of the mutual friction and the normal fluid shear viscosity.
Finally, we note that larger wave vector oscillations corresponding
to periodic motions on shorter length scales may lead to phenomena that
are responsible for the observed timing noise in pulsars.

Our model necessarily involves certain approximations. For instance, we have adopted the two-fluid superfluid hydrodynamics, which should be modified in order to account for the multiple fluids in the neutron star's core~\cite{MFLUIDS1,MFLUIDS2,MFLUIDS3,Mendell:1997fu,MFLUIDS4,Sedrakian:1998ki,MFLUIDS5}.
Furthermore, the cylindrical symmetry of our setup and the assumption of uniform density need to be relaxed in more realistic treatments of spherical superfluid shells with density gradients.

\section*{ACKNOWLEDGMENTS}

J.N. acknowledges support by the
Frankfurt International Graduate School for Science (FIGSS).

\end{document}